\documentclass[twoside,a4paper]{article}
\usepackage{fleqn,epsf,graphicx}
\usepackage{authblk}
\usepackage{hyperref}
\usepackage{cite}
\usepackage{tikz}
\usetikzlibrary{arrows,calc, shapes, decorations.pathreplacing}

\begin{document}

\title{Analysis framework for the J-PET scanner}

\author{W.~Krzemie\'n$^{a}$, A.~Gajos$^{b}$, A.~Gruntowski$^{b}$, K.~Stola$^{b}$, D.~Trybek$^{b}$, T.~Bednarski$^{b}$, P.~Bia\l as$^{b}$, E.~Czerwi\'nski$^{b}$, D.~Kami\'nska$^{b}$, \L .~Kap\l on$^{b,c}$, A.~Kochanowski$^{d}$, G.~Korcyl$^{b}$, J.~Kowal$^{b}$, P.~Kowalski$^{e}$, T.~Kozik$^{b}$, E.~Kubicz$^{b}$, P.~Moskal$^{b}$, Sz.~Nied\'zwiecki$^{b}$,  M.~Pa\l ka$^{b}$, L.~Raczy\'nski$^{e}$, Z.~Rudy$^{b}$, P.~Salabura$^{b}$, N.G.~Sharma$^{b}$, M.~Silarski$^{b}$, A.~S\l omski$^{b}$, J.~Smyrski$^{b}$, A.~Strzelecki$^{b}$, A.~Wieczorek$^{b,c}$, W.~Wi\'slicki$^{e}$, M.~Zieli\'nski$^{b}$, N.~Zo\'n$^{b}$}

\affil{
       
       $^{a}$High Energy Physics Division, National Centre for Nuclear Research, 05-400 Otwock-\'Swierk, Poland\\
       $^{b}$
  Faculty of Physics, Astronomy and Applied Computer Science, Jagiellonian University, 30-348 Cracow, Poland\\
       $^{c}$Institute of Metallurgy and Materials Science of Polish Academy of Sciences, 30-059 Cracow, Poland\\
       $^{d}$Faculty of Chemistry, Jagiellonian University, 30-060 Cracow, Poland\\
       $^{e}$\'Swierk Computing Centre, National Centre for Nuclear Research, 05-400 Otwock-\'Swierk, Poland\\
     }

\maketitle                   


\begin{abstract}
  J-PET analysis framework is a flexible, lightweight, ROOT-based software package 
  which provides the tools to develop reconstruction and calibration procedures for PET tomography. In this article we present the implementation of the full data-processing chain in the J-PET framework which is used for the data analysis of the J-PET tomography scanner. The Framework incorporates automated handling of PET setup parameters' database as well as high level tools for building data reconstruction procedures. Each of these components is briefly discussed.
\end{abstract}

\section{Introduction}

Positron Emission Tomography (PET) is one of the most accurate, non-invasive techniques used in modern medical diagnostics
to produce the metabolic image of patient by reconstructing the spatial distribution 
of a radiopharmaceutical administered into the patient's blood. The signals coming from the patient's body in the form of gamma quanta are recorded by scintillator-based tomography scanners.

The Jagiellonian PET (J-PET) collaboration develops a novel concept of a Time-of-Flight PET (TOF-PET) apparatus based on plastic scintillators, which largely improve the field of view and the timing resolution of the TOF method \cite{PCT2010,JPET-Genewa,NovelDetectorSystems,StripPETconcept,TOFPETDetector}. In order to achieve this goal, the J-PET detector consists of large strips of plastic scintillators read out at both ends by photomultipliers \cite{Moskal:2014sra}. The photomultiplier signals are sampled and recorded by dedicated Front-End Electronics (FEE) specifically designed for this project \cite{palka, korcyl}.
  
\section{Technology}
J-PET analysis framework has been developed mainly in the C++ programming language applying the object-oriented approach. Additionally, small parts of the code are written in Python and Perl. The C++11 standard and Boost Libraries have been used. Boost Unit Test Framework~\cite{Boost} has been chosen to construct a set of automatic test to assure the quality of the code.
The package is based on the Open Source library ROOT~\cite{Brun:1997pa} which is a de facto standard  in most of the modern high-energy physics and hadron physics experiments.
The documentation of the code is automatically generated using dedicated Doxygen program~\cite{doxygen}. 
Besides, several external libraries are used: tinyxml~\cite{tinyxml} and libconfig++~\cite{libconfig}.
The database handler module responsible for the communication with the PostgreSQL database uses PostgreSQL Application Programming Interface (API)~\cite{pqxx}. The SQL queries have been enclosed in the PostgreSQL functions which are remotely called from the framework code. 

\section{Framework architecture}

The basic architecture of the J-PET framework was already explained 
in ref.~\cite{krzemien}. In this section we shortly review the main idea and explain in more detail the integration of the framework with the parametric database.

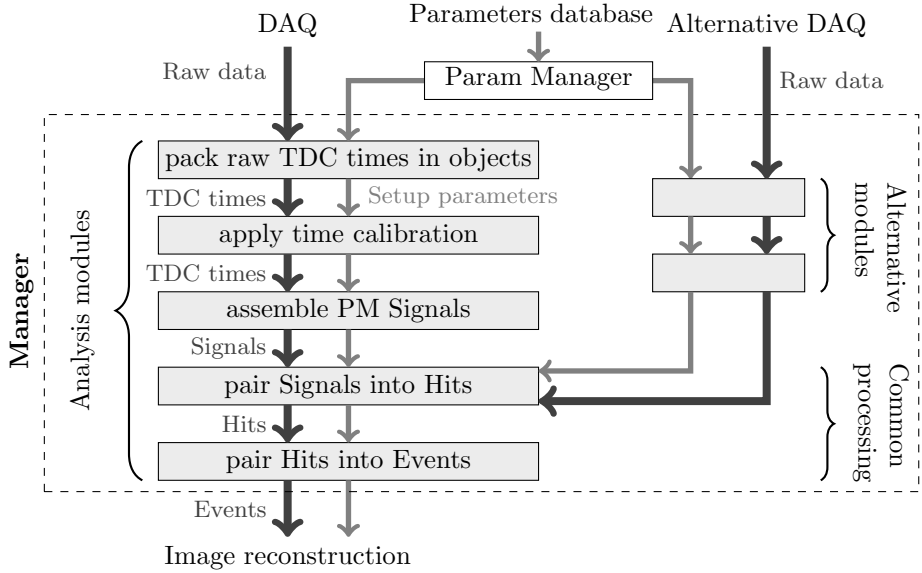
\begin{figure}[ht]
  \centering
  \begin{tikzpicture}[
  scale = 1,
  ]
  \tikzstyle{data arrow} = [line width=3pt, darkgray]
  \tikzstyle{par arrow}  = [line width=2pt, gray]
  \tikzstyle{module}  = [fill=lightgray!30!white]
  \coordinate (A) at (-3, 1);
  \coordinate (B) at (-3, 0);
  \coordinate (C) at (-3, -1);
  \coordinate (D) at (-3, -2);
  \coordinate (E) at (-3, -3);
  \coordinate (DAQ) at ($(A)+(-0.8,1.5)$);
  \coordinate (PM) at (-2, 1.8);
  \coordinate (DB) at (-2, 2.7);
  \coordinate (A2) at (2, 0.5);
  \coordinate (B2) at (2, -0.5);
  \coordinate (DAQ2) at ($(A2)+(0.5,2)$);
  %
  \draw[module] ($(A)-(2.5,0.25)$) rectangle ($(A)+(2.5,0.25)$);
  \node[] at (A) {pack raw TDC times in objects};
  \draw[module] ($(B)-(2.5,0.25)$) rectangle ($(B)+(2.5,0.25)$);
  \node[] at (B) {apply time calibration};
  \draw[module] ($(C)-(2.5,0.25)$) rectangle ($(C)+(2.5,0.25)$);
  \node[] at (C) {assemble PM Signals};
  \draw[module] ($(D)-(2.5,0.25)$) rectangle ($(D)+(2.5,0.25)$);
  \node[] at (D) {pair Signals into Hits};
  \draw[module] ($(E)-(2.5,0.25)$) rectangle ($(E)+(2.5,0.25)$);
  \node[] at (E) {pair Hits into Events};
  \draw[decorate,decoration={brace,mirror,raise=6pt,amplitude=10pt}, thick] ($(A)+(-2.5,0.25)$)--($(E)-(2.5,0.25)$);
  \node[rotate=90] at ($(C)-(3.5,0)$) {Analysis modules};
  \draw[data arrow,->] ($(A)+(-0.8,-0.25)$) -- ($(B)+(-0.8,+0.25)$) node[midway,left] {\small TDC times\ \ };
  \draw[data arrow,->] ($(B)+(-0.8,-0.25)$) -- ($(C)+(-0.8,+0.25)$) node[midway,left] {\small TDC times\ \ };
  \draw[data arrow,->] ($(C)+(-0.8,-0.25)$) -- ($(D)+(-0.8,+0.25)$) node[midway,left] {\small Signals\ \ };
  \draw[data arrow,->] ($(D)+(-0.8,-0.25)$) -- ($(E)+(-0.8,+0.25)$) node[midway,left] {\small Hits\ \ };
  \draw[data arrow,->] ($(E)+(-0.8,-0.25)$) -- ($(E)+(-0.8,-1)$) node[midway,left] {\small Events\ \ };
  \draw[par arrow,->] ($(A)+(0,-0.25)$) -- ($(B)+(0,+0.25)$) node[midway,right] {\small \ Setup parameters};
  \draw[par arrow,->] ($(B)+(0,-0.25)$) -- ($(C)+(0,+0.25)$) node[midway,right] {};
  \draw[par arrow,->] ($(C)+(0,-0.25)$) -- ($(D)+(0,+0.25)$) node[midway,right] {};
  \draw[par arrow,->] ($(D)+(0,-0.25)$) -- ($(E)+(0,+0.25)$) node[midway,right] {};
  \draw[par arrow,->] ($(E)+(0,-0.25)$) -- ($(E)+(0,-1)$) node[midway,right] {};
  \draw[data arrow,->] (DAQ) -- ($(A)+(-0.8,+0.25)$) node[near start,left] {\small Raw data\ \ };
  \node[above] at (DAQ) {DAQ};
  \draw[dashed] (-7, -3.4) rectangle (4.5,1.6);
  \node[rotate=90] at (-7.3,-1) {\bf Manager};
  \draw[par arrow,->] ($(PM)+(0,0.25)$) -| ($(A)+(0,0.25)$);
  \draw[par arrow,->] ($(PM)+(3,0.25)$) -| ($(A2)+(-0.5,0.25)$);
  \draw[] (PM) rectangle ($(PM)+(3,0.5)$);
  \node[right, above] at ($(PM)+(1.5,0)$) {Param Manager};
  \draw[par arrow,->] ($(DB)+(1.5,0)$) -- ($(PM)+(1.5,0.5)$);
  \node[right, above] at ($(DB)+(1.4,0)$) {Parameters database};
  \draw[module] ($(A2)-(1,0.25)$) rectangle ($(A2)+(1,0.25)$);
  \draw[module] ($(B2)-(1,0.25)$) rectangle ($(B2)+(1,0.25)$);
  \draw[par arrow,->] ($(A2)+(-0.5,-0.25)$) -- ($(B2)+(-0.5,+0.25)$) node[midway,left] {};
  \draw[data arrow,->] ($(A2)+(0.5,-0.25)$) -- ($(B2)+(0.5,+0.25)$) node[midway,right] {};
  \draw[par arrow,->] ($(B2)+(-0.5,-0.25)$) |- ($(D)+(2.5,+0.2)$) node[midway,left] {};
  \draw[data arrow,->] ($(B2)+(0.5,-0.25)$) |- ($(D)+(2.5,-0.2)$) node[midway,right] {};
  \draw[data arrow,->] (DAQ2) -- ($(A2)+(0.5,+0.25)$) node[near start,right] {\small Raw data};
  \node[above] at (DAQ2) {Alternative DAQ};
  \draw[decorate,decoration={brace,raise=6pt,amplitude=5pt}, thick] ($(A2)+(1,0.25)$)--($(B2)+(1,-0.25)$);
  \node[text width = 1.4cm,rotate=-90] at ($(B2)+(2,0.5)$) {Alternative modules};
  \draw[decorate,decoration={brace,raise=6pt,amplitude=5pt}, thick] ($(D)+(6,0.25)$)--($(E)+(6,-0.25)$);
  \node[text width = 1.6cm,rotate=-90] at ($(B2)+(2,-2)$) {Common processing};
  \node[below] at ($(E)+(-0.8,-1)$) {Image reconstruction};

\end{tikzpicture}
  \caption{\label{scheme} Scheme of an example data processing chain with the J-PET Framework.}
  \label{fig:data-proc}
\end{figure}

The data reconstruction in modern PET tomography is a complex multi-stage process including several low and high level reconstruction algorithms and calibration procedures, which needs a
special computing support~\cite{Wislicki}. Also, some
monitoring procedures are required to be implemented.
The basic concept of the J-PET Framework is the decomposition of analysis chains into series of standardized modular blocks. 
Each module corresponds to a particular computing task, e.g. reconstruction algorithm or calibration procedure,  with a defined input and output methods. 
The processing chain is built by registration of chosen modules in the JPetManager, which is responsible for the synchronization of the data flow between the modules. 
This approach has several advantage e.g. it permits to quickly interchange modules and to create a processing chain that can be used for different experimental setups, e.g. in the J-PET project, the substitution of the lowest level processing modules allows to analyse both the raw data delivered by the full Data Acquisition system (DAQ) but also data registered in the diagnostic setup using a digital oscilloscope (see Fig.~\ref{scheme}).

In addition, the framework is well suited to develop procedures in parallel, by taking advantage of the standardize module structure and predefined input/output. This feature is particularly important in the prototype phase of the project, where the several reconstruction procedures are developed and tested at the same time by different subgroups.

\subsection{Data processing}
The J-PET Analysis framework aims at providing the possibility to implement reconstruction procedures to users even with little programming experience through a convenient and simple API abstracting the technical details. 
Creation of a simple analysis stage requires providing a module class which basically comprises three methods only: 
one to be called prior to data processing, 
one after processing, and a method called for each event in the dataset. 
Any non-standard processing can be achieved, however, simply by overwriting more specialized methods inherited from the base module. 
The elementary pieces of recorded data are enclosed in objects containing not only the experimental and reconstructed information such as recorded time signals from Time-to-Digital Converters (TDC), but also pointers to objects representing parameters of all the elements of the DAQ system related to each signal. 
These pointers are implemented using the {\emph{TRef}} mechanism from the ROOT framework which allows them to be persistently stored and retrieved from ROOT files.  
Therefore, the set of parameters from the database is downloaded only once before the processing and then stored along with data files which allows to access each necessary parameters related to a certain signal with $\mathcal{O}(1)$ operations at any time.
For every class a list of handful methods defines an interface that makes the access to data easy.
The relation between the parametric objects reflexes the logical or physical connection between the objects they represent, e.g. JPetPM object describing the proprieties and setting of a photomultiplier, points to JPetScint object, which corresponds to a scintillator module to which this photomultiplier is connected. The information about all the relations is encoded in the parametric database and then is restored while generating parameter objects.

A typical processing scheme with the Framework is split between several modules,  each analysing a set of data objects and combining them into more complex  structures as shown in Fig.~\ref{fig:data-proc}. The analysis starts with a set of raw TDC signals and uses them to reconstruct shapes of electric signals from photomultipliers.  Next, pairs of signals corresponding to a single photon hit in a scintillator strip are identified and used to build hit objects.  Finally,  pairs of hits from a single annihilation event are merged into Event structures which in turn constitute the input to image reconstruction procedures designed for the J-PET \cite{Slomski,Bialas1,Bialas2}.  Each of these steps is performed by a separate module which allows for easy exchange of the procedures in case of change in the DAQ system. For instance,  tests described in~\cite{Moskal:2014sra} were performed with a Serial Data Analyzer providing dense signal sampling with the dedicated DAQ, but the respective data can be processed by the same framework with an exchange of the lower-level modules while the higher-level processing modules are common for all DAQ chains (see Fig.~\ref{fig:data-proc}).

\subsection{Connection with the parametric database}
During the PET examination all information about detector setup and experimental conditions must be stored. In the J-PET project the dedicated parametric database is used for this purpose~\cite{Czerwinski:2014awa}. 
It uses PostgreSQL object-relational database management system. 
The database contains records about hardware settings, calibration constants, 
initial setup parameters, alignment of the detectors and software configurations. 
For each single measurement all these settings must be saved and available at any time for analysis and data processing.
The communication between the parametric database and the analysis framework is done 
by a dedicated module (DBHandler)~\cite{Gruntowski_MSC}. 
It works as an interface handling all connection operations and hiding the details of the database structure from users. 
The Parameter Manager  generates necessary information for analysis and reconstruction modules, 
based on the data provided by the DBHandler, 
in the form of collections of parametric objects describing geometry, 
scintillators, photomultipliers, voltage etc. 
The Parameter Manager is also responsible for saving and reading the subsets of parameter objects in ASCII or ROOT file formats. 
The information retrieved from the database is stamped with the run number, which uniquely determines a given measurement condition set.  
The whole process is fully automated. The user has to give a valid run number identifier and the appropriate set of parameters will be automatically created and loaded in its analysis module.

\section{Summary}
In this article we have presented a software framework for the PET tomography data analysis and reconstruction, that has been developed 
as a part of the J-PET project. 
The package is ROOT-based and it is inspired by several solutions from the hadronic and
high energy physics experiments~\cite{wasa,kloe2,cosy11,hades}.
The JPet Framework provides a set of programming tools and predefined data objects 
that helps in the implementation process even for unexperienced users.
As it was shown in the examples, the proposed modular architecture 
permits to efficiently handle different data sources, DAQ setups and detector 
geometries by exchanging the low-level modules in the processing chain.
Finally, the J-PET Framework can provide information on the measurement conditions, geometries and hardware setups
as a set of parameter objects generated based on the database registers in a fully automated way.
The first full analysis chain starting from the "raw" data delivered by scanner data acquisition system up to the reconstructed set of
Line of Response has been implemented. Currently, the full signal reconstruction in scintillator methods based on the Compressive Sampling~\cite{Raczynski:2014poa} and the library of 
signals~\cite{zon} are being incorporated.

\section{Acknowledgements}
We acknowledge technical and administrative support by T. Gucwa-Rys, A. Heczko,
M. Kajetanowicz, G. Konopka-Cupiał, W. Migda\l, and the financial support by the Polish National Center for Development and Research through grant No. INNOTECH-K1/IN1/64/159174/NCBR/12, the Foundation for Polish Science through MPD programme and the EU, MSHE Grant No. POIG.02.03.00-161 00-013/09 and Doctus – the Malopolska PhD Scholarship Fund.
%
%

\end{document}